\documentclass[%
 reprint,
 amsmath,amssymb,
 aps,
]{revtex4-1}

\usepackage{graphicx}
\usepackage{dcolumn}
\usepackage{bm}
\usepackage{dsfont}

\usepackage{color}
\usepackage{ulem}

\def\s{S^{\rm QCD}}

\def\m{M^{\rm QCD}}

\def\spi{{S'}^{\rm\,\, QCD}}

\begin{document}

\preprint{APS/123-QED}

\title{Connections between chiral Lagrangians and QCD sum-rules
}
\author{Amir H. Fariborz}
\email{fariboa@sunyit.edu}
\affiliation{Department of Mathematics/Physics,  SUNY Polytechnic Institute, Utica, NY 13502, U.S.A.
}
\author{A.~Pokraka}
\email{amp348@mail.usask.ca}
\affiliation{%
Department of Physics and
Engineering Physics, University of Saskatchewan, Saskatoon, SK,
S7N 5E2, Canada
}%
\author{T.G. Steele}
\email{Tom.Steele@usask.ca}
\affiliation{%
Department of Physics and
Engineering Physics, University of Saskatchewan, Saskatoon, SK,
S7N 5E2, Canada
}%

\begin{abstract}
It is shown how a chiral Lagrangian framework
can be used to
derive  relationships  connecting quark-level QCD correlation functions  to  mesonic-level two-point functions.
Crucial ingredients of this connection are scale factor matrices relating each distinct quark-level  substructure (e.g., quark-antiquark, four-quark) to its mesonic counterpart.
 The scale factors and  mixing angles are combined into a projection matrix to obtain the physical (hadronic) projection of the QCD correlation function matrix.
 Such relationships provide a powerful bridge between chiral Lagrangians and QCD sum-rules that are particularly effective in studies of the substructure of light scalar mesons with multiple complicated resonance shapes and substantial underlying mixings.
The validity of these connections is demonstrated for the example of the isotriplet $a_0(980)$-$a_0(1450)$ system, resulting in an unambiguous determination of the scale factors from the combined inputs of QCD sum-rules and chiral Lagrangians.   These scale factors lead to a remarkable  agreement between the quark condensates in QCD and the mesonic vacuum expectation values
that induce spontaneous chiral symmetry breaking in chiral Lagrangians.
This concrete example  shows a clear sensitivity to the underlying $a_0$-system mixing angle, illustrating the value of this methodology in extensions to more complicated mesonic systems.
\end{abstract}
\maketitle

Scalars are the Higgs bosons of QCD and induce spontaneous chiral symmetry breaking and their internal structure holds key information about  nonperturbative QCD.   However,  understanding the light scalar sector of QCD has turned out to be considerably more challenging than other sectors \cite{PDG,Weinberg_13,07_KZ}.
Particularly the investigation of the quark substructure of scalar mesons continues to pose various complications, mostly stemming from the underlying mixings among different quark and glue components.    On the one hand,  the experimental data on light scalars,  which originate from various low-energy decays and scatterings, seems to be less certain than other better understood channels such as vectors or pseudoscalars, while on the other hand,  the theoretical foundation for understanding scalars faces  challenges within the quark model.
The theoretical challenges  include describing the light and inverted scalar mass spectrum   below 1 GeV,  which finds a fundamental explanation in terms of four-quark states described by the MIT bag model \cite{Jaf}, as well as describing the existing deviations from a quark-antiquark substructure of scalar states above 1 GeV,  which finds a natural explanation in an underlying mixing among scalars below and above 1 GeV \cite{Mec} (other works on mixing  include Refs.~\cite{close,mixing,NR04,06_F,08_tHooft,global,07_FJS4,05_FJS}).
QCD sum-rules also find evidence of an inverted spectrum for four-quark states \cite{Zhang:2000db,Brito:2004tv,Chen:2007xr}.

Such mixing patterns and global relations are far from the focus of chiral perturbation theory \cite{ChPT,01_CGL,06_CCL,06_Pelaez,OOP} which aims to probe the near-threshold region in a systematic way.  Those formulations of  chiral Lagrangians (both linear \cite{NR04,global,07_FJS4,05_FJS} as well as nonlinear models \cite{Mec,06_F,SS,BFSS2,Blk_rad,RCPT}) that explicitly include scalar resonances have been shown to be effective frameworks for understanding scalars.
In this approach, the framework is developed  on the basis of chiral symmetry,  the mechanisms for its breakdown, and the model parameters are found from fits to the available experimental data on scalars below 2 GeV.      Particularly, it is seen that the underlying mixing patterns among scalar states provides an indirect  probe into their substructure \cite{BFSS2,06_F,global}.    It is shown that the scalars below and above 1 GeV have considerable mixings,   with those below 1 GeV being  predominantly of four-quark nature whereas those above 1 GeV being closer to quark-antiquarks. In addition,  two of them, $f_0(1500)$ and $f_0(1710)$, tend to acquire a substantial  admixtures of glue \cite{close,06_F,oller}.  QCD sum-rules have also provided insight into these gluonic mixings
(see e.g., \cite{qcdsr_glue_mix1,qcdsr_glue_mix2,qcdsr_glue_mix3,qcdsr_glue_mix4,qcdsr_glue_mix5,GSR_qq_results_mix} and review of earlier results in Ref.~\cite{narison_review}), as well as other methods \cite{other_mix1,other_mix2}.

While the probe of quark substructure of scalars in the chiral Lagrangian approach is robust and coherent, it is  indirect and limited.   One of the limitations is the fact that   the four-quark chiral nonet can be mapped to different quark compositions, each   representing different flavor, spin and color configurations (e.g., ``molecular'' type,  diquark-antidiquark type).  Since  these  different compositions transform in exactly the same way under chiral transformations,     the chiral Lagrangians become oblivious to  differences in internal quark substructure of the four-quark chiral nonet.   In Refs.~\cite{global,07_FJS4,05_FJS}, as discussed below,  it is shown that   ${\rm U}(1)_{\rm A}$  differentiates between two-quark and four-quark nonets, but not between different types of four-quark substructures.
It is therefore desirable to supplement these investigations with a method that directly probes the quark substructure.  This limitation of chiral Lagrangians can be in principle addressed by a linkage to QCD sum-rules.

Conversely, QCD sum-rules can benefit from chiral Lagrangians.
  QCD sum-rules exploit quark hadron duality and connect  fundamental QCD to hadronic physics \cite{SVZ}.  For  probing a light, broad and highly overlapping state, such as the sigma meson,  a simple Breit-Wigner shape may not be a realistic parametrization.   In addition,  when there are complicated underlying mixings,  a realistic modeling of the hadronic side in QCD sum-rules that includes the effects of such underlying mixing phenomena seems   necessary.    Chiral Lagrangians (such as those of Refs.~\cite{global,07_FJS4,05_FJS}) where their model parameters have been fixed by various low-energy data  can provide important information on both the resonance shapes and mixings.

The main purpose of this paper is to demonstrate how
 chiral Lagrangians and QCD sum-rules can play complementary roles in
forming a bridge that ultimately connects the relevant low-energy experimental data to fundamental QCD.
Key parts of this bridge are scale factor matrices  relating each distinct quark-level  substructure  to its chiral Lagrangian mesonic counterpart.   The scale factors and  mixing angles are combined into a projection matrix to obtain the physical (hadronic) projection of the QCD correlation function matrix.
This bridge is therefore mutually beneficial: Chiral Lagrangians provide phenomenological information to guide QCD sum-rules,
 and QCD sum-rules can provide additional methodologies for distinguishing between different chiral Lagrangian scenarios, deepening our understanding of the
chiral Lagrangian approach.

Although this work is motivated by the scalar channel,   it is extendable to other channels such as the pseudoscalar and vector channels.   Here, we use the generalized linear sigma model of Ref.~\cite{global} to demonstrate how  connections can be made with QCD sum-rules, but the method is more general and can be extended to nonlinear chiral Lagrangians such as those in
Refs.~\cite{BFSS2,Mec}.    Also,  our focus in this work is the relatively less-problematic $I=1$ channel that allows us to demonstrate the connection without complicating factors.
We intend to extend the present work to  nonlinear chiral Lagrangians as well as other channels ($I=1/2$ and $I=0$)  in future works.

We begin by defining our notation. At the mesonic-level,   we employ the generalized linear sigma model of Refs.~\cite{global,07_FJS4,05_FJS} which is formulated in terms of two chiral nonets  $M$ and $M'$ that respectively represent a quark-antiquark nonet and a four-quark nonet (a ``molecule'' type and/or a diquark-antidiquark type) underlying substructure.      Both chiral nonets transform in the same way under chiral transformations but differently under U(1)$_{\rm A}$:
\begin{eqnarray}
M & \rightarrow& U_L \,  M \,  U_R^\dagger\,,  \qquad M\rightarrow e^{2i\nu}M\nonumber \\
M' & \rightarrow& U_L \,  M' \,  U_R^\dagger\,, ~\quad  M'\rightarrow e^{-4i\nu}M'~.
\label{M_trans}
\end{eqnarray}
The axial charge is the main tool for distinguishing these two nonets.   Each of these two chiral nonets can be expressed in terms of a
scalar and a pseudoscalar meson nonet
\begin{eqnarray}
M & = & S + i\phi \nonumber \\
M' & = & S' + i \phi'
\label{M_SP}
\end{eqnarray}
where the two scalar meson nonets contain the two- and four-quark ``bare'' (unmixed) scalars
\begin{equation}
S=
\begin{pmatrix}
S_1^1 & a_0^+ & \kappa^+  \\
a_0^- & S_2^2 & \kappa^0 \\
\kappa^- & {\bar \kappa}^0 & S_3^3
\end{pmatrix},
\hskip 0.5cm
S' =
\begin{pmatrix}
{S'}_1^1 & {a'}_0^+ & {\kappa'}^+  \\
{a'}_0^- & {S'}_2^2 & {\kappa'}^0 \\
{\kappa'}^- & {\bar {\kappa'}}^0 & {S'}_3^3 \\
\end{pmatrix}
\label{SpMES}
\end{equation}
and similar matrices for $\phi$ and $\phi'$.   The framework of Ref.~\cite{global} provides a detailed analysis of the mixing between these two ``bare'' nonets and how that results in a description of mass spectrum, decay widths and scattering analysis of  scalar as well as pseudoscalar mesons below and above 1 GeV.    In this picture,  the physical isotriplet scalars become a linear admixture of two- and  four-quark components
\begin{equation}
\begin{pmatrix}
a_0^0(980)\\
a_0^0(1450)
\end{pmatrix}
=
L_a^{-1}
\begin{pmatrix}
\left(S_1^1 - S_2^2\right)/\sqrt{2}\\
\left({S'}_1^1 - {S'}_2^2\right)/\sqrt{2}
\end{pmatrix}
\end{equation}
where $L_a^{-1}$ is a 2$\times$2 rotation matrix (here we use the matrix computed in \cite{global}).

The transformation properties (\ref{M_trans}) as well as decompositions (\ref{M_SP}) are direct consequences of the assumed underlying quark configurations.   The two mesonic-level  chiral nonets  $M$ and $M'$ can be mapped to the quark-level chiral nonets $\m$ and $M'^{\,\rm QCD}$.   For example, Eq.~(\ref{M_trans}) implies
\begin{equation}
(\m)_a^b \propto ({\bar q}_R)^b ({q_L})_a  \Rightarrow \left(S^{\rm QCD}\right)_a^b \propto q_a {\bar q}^b~.
\label{SQCD}
\end{equation}
To make the connection to the quark world we need to make a specific choice for the proportionality factor, where here we choose $\left(S^{\rm QCD}\right)_a^b  = q_a {\bar q}^b$ (there is no loss of generality or physical consequences associated with this choice).
The presence of a composite operator in \eqref{SQCD} foreshadows our connection to QCD sum-rules because they are based upon correlation functions
of composite operators \cite{SVZ}.
 Similarly,  $M'^{\,\rm QCD}$ can be  mapped to quark-level composite field configurations.   However, in this case there are several options,  each representing a different angular momentum, spin, flavor and color configurations for diquark-antiquark combination.   Here we do not list such quark configurations and only give the specific form below that we have used for our example analysis.

We assume a direct relationship between $M^{\rm QCD}$ and $M$ via a scale matrix  $I_{M}$ that adjusts the mass dimensions
\begin{equation}
M=I_{M} M^{\rm QCD}~.
\label{M_scale}
\end{equation}
Eqs.~\eqref{SQCD} and \eqref{M_scale} imply that
under chiral transformations, the physical chiral nonet $M$ transforms  as follows \begin{equation}
{(I_{M})}^{-1} M \rightarrow U_L {(I_{M})}^{-1} M U_R^\dagger
\end{equation}
where by multiplying both sides  by $I_{M}$ from left
\begin{equation}
M \rightarrow I_{M} U_L {(I_{M})}^{-1} M U_R^\dagger~.
\end{equation}
If $M$ is  to transform in the same way as (\ref{M_trans}), then
\begin{equation}
I_{M}  U_L I^{-1}_{M}=U_L \hskip 0.5cm \Rightarrow \hskip 0.5cm [U_L,I_{M}]=0~.
\end{equation}
Similarly it can be shown (based on chiral tranformation of $M^\dagger$)
\begin{equation}
[U_R,I_{M}]=0
\end{equation}
which means $I_{M}$ is a multiple of unit matrix, $I_M=-{m_q\over \Lambda^3}\times \mathds{1}$, where $\Lambda$ (that must be determined) is a scale factor connecting mesonic and QCD fields.
The quark mass factor $m_q=(m_u+m_d)/2$ will ultimately result in renormalization-group invariant currents as discussed below.
Similarly,
\begin{equation}
M'=I_{M'} M'^{\,\rm QCD}
\label{Mp_scale}
\end{equation}
and we can show that $I_{M'}$ is a multiple of unit matrix, i.e.
$I_{M'}={1\over {{\Lambda'}^5}}\times  \mathds{1}$, where $\Lambda'$ is the scale factor for the $M'$ structures (similar to $\Lambda$, the scale factor $\Lambda'$ must also be determined).

At this point the general methodology should be evident: each distinct substructure
(e.g., two-quark versus four-quark)
 requires a separate scale factor connecting the meson fields to the QCD currents.
Methods for determining these scale factors are developed below, and are a central feature of our analysis.
To examine the validity of the connections  between chiral Lagrangians and QCD sum-rules, we consider a specific example of
the isotriplets $a_0(980)$ and $a_0(1450)$, for which
\begin{equation}
\begin{split}
{\bf A}=&
\begin{pmatrix}
a_0^0(980)\\
a_0^0(1450)
\end{pmatrix}
= L_a^{-1}
\begin{pmatrix}
\left(S_1^1 - S_2^2\right)/\sqrt{2}\\
\left({S'}_1^1 - {S'}_2^2\right)/\sqrt{2}
\end{pmatrix}
\\
=&
\frac{L_a^{-1} I_a}{\sqrt{2}}
\begin{pmatrix}
        \left( S^{\rm QCD} \right)_1^1 -
       \left( S^{\rm QCD} \right)_2^2
       \\[5pt]
\left( \spi \right)_1^1 - \left(\spi\right)_2^2
\end{pmatrix}
\end{split}
\end{equation}
where $L_a$ is the rotation matrix for isovectors defined in \cite{global}
and $I_a$    is formed out of the scale factors defined for the two chiral nonets in (\ref{M_scale}) and (\ref{Mp_scale})
\begin{equation}
L^{-1}_a=\begin{pmatrix}
\cos\theta_a & -\sin\theta_a
\\
\sin\theta_a & \cos\theta_a
\end{pmatrix}
\,,~I_a =
\begin{pmatrix}
{-m_q\over \Lambda^3} &0 \\
0 & {1\over {{\Lambda'}^5}}
\end{pmatrix}
~.
\end{equation}

We can now associate the QCD-level fields $\s$ and $\spi$ with similar-structure currents  that are needed for  QCD sum-rule techniques, so that
  the physical currents can be parameterized in the same way as that of states:
\begin{equation}
J^{\rm QCD} =\frac{1}{\sqrt{2}}
\begin{pmatrix}
        \left( S^{\rm QCD} \right)_1^1 -
       \left( S^{\rm QCD} \right)_2^2
       \\[5pt]
\left( \spi \right)_1^1 - \left(\spi\right)_2^2
\end{pmatrix}
\end{equation}
and this is related to the projected physical currents
\begin{equation}
J^P = L_a^{-1} I_a J^{\rm QCD}
\end{equation}
that probe the physical particles through correlation functions
\begin{equation}
\Pi^{\rm QCD}_{mn}(x) =\langle 0| {\rm T}  \left[ J^{\rm QCD}_m (x) J^{\rm QCD}_n(0) \right] |0 \rangle~.
\end{equation}

Then the projected physical  correlators can be defined
\begin{equation}
\begin{split}
\Pi^P_{ij}(x)=&\langle 0| {\rm T} \left[ J^P_i (x) J^P_j(0) \right] |0 \rangle
\\
=& \left(L_a^{-1}\right)_{il}\left(I_a\right)_{lm} \Pi^{\rm QCD}_{mn}(x) \left(I_a\right)_{nk} \left(L_a\right)_{kj}
~.
\end{split}
\end{equation}
We thus obtain (after a Fourier transform to momentum space)  the projected physical QCD correlation function matrix
\begin{equation}
 \Pi^P(Q^2) = {\widetilde {\cal T}}^a \Pi^{\rm QCD}(Q^2)  {\cal T}^a\,,~~{\cal T}^a= I_a \, L_a
\label{PiP}
\end{equation}
where ${\widetilde {\cal T}^a}$  denotes the transpose of the matrix ${\cal T}^a$ and $Q^2=-q^2$.
Although \eqref{PiP} has been developed for the specific case of the $a_0$ system, its form is quite general and can be easily extended to other systems by augmenting the matrix $I_a$ to include a scale factor for each substructure and expanding the mixing  matrix $L_a$ accordingly.  The key point is  that a matrix ${\cal T}^a$, composed of scale factors and mixing angles, can be used to obtain the physical projection of  the QCD correlation function matrix.

On the other hand,  the  hadronic contribution to the physical correlator
is determined via  the mesonic fields and a QCD continuum
\begin{equation}
\begin{split}
\Pi^{\rm H}_{ij} \left(q^2\right)&=\int d^4x\, e^{iq\cdot x}
\langle 0| {\rm T} \left[ {\bf A}_i (x) {\bf A}_j(0) \right] |0 \rangle
\\
&=\delta_{ij} \,
\left(
{1\over {m_{ai}^2 - q^2 -i m_{ai}\Gamma_{ai}}}
+ {\rm cont.}\right)~.
\end{split}
\label{PiH}
\end{equation}
The effect of final-state interactions in the $\pi\eta$ channel is not as significant as this effect in the $\pi\pi$ ($\pi K$) channel in which the sigma (kappa) is probed, and therefore in the first approximation is neglected here.
The last term represents the QCD continuum contribution inherent in QCD sum-rule methods \cite{SVZ}.
This hadronic correlator is related to the projected physical QCD correlator (\ref{PiP}) by a standard dispersion relation.

Eqs.~\eqref{PiP}, (\ref{PiH}) can be used to express some constraints on  the correlators.     In the isotriplet channel case the vanishing of the off-diagonal elements of $\Pi^P$ (which is a 2 $\times$ 2 matrix in this case)
leads to the following self-consistency condition for
$\Pi_{12}^{\rm QCD}$:
\begin{equation}
\Pi_{12}^{\rm QCD} = -
\left[
{
 {    {\widetilde {\cal T}}^a_{11} \Pi_{11}^{\rm QCD} {\cal T}^a_{12}
    + {\widetilde {\cal T}}^a_{12} \Pi_{22}^{\rm QCD} {\cal T}^a_{22}
 }
 \over
 {{\widetilde {\cal T}}^a_{11}  {\cal T}^a_{22} + {\widetilde {\cal T}}^a_{12}  {\cal T}^a_{22}
 }
}
\right]~.
\label{G12}
\end{equation}
This expression can be used in various ways.  If the off-diagonal QCD correlator is known, then this expression can be used to assess the self-consistency of (or to augment) the analysis. Alternatively, because off-diagonal correlators typically require higher-loop calculations  and may be unknown (and impractical to calculate), this expression can be used to eliminate the off-diagonal correlator.  This latter approach will be used in our analysis of the $a_0$ system.

Following standard QCD sum-rule methodologies, an operator (e.g., Borel transformation operator) is applied to the dispersion relation connecting the QCD and hadronic contributions to the projected physical correlators \cite{SVZ}. Because the Laplace QCD sum-rules emphasize the lowest state, they are not ideally suited to our purposes because they may obscure the mixing phenomena central to our considerations. We therefore focus on the lowest-weight Gaussian sum-rules that provides similar weight to ground and excited states \cite{gauss,harnett_quark}
\begin{equation}
G_0^{\rm P} ({\hat s}, \tau, s_0, s_{th}) =\frac{1}{\sqrt{4\pi\tau}}
\int\limits_{s_{th}}^{s_0}   {\rm exp} \left[  {{-({\hat s} - t)^2}\over {4\tau}}\right]\, {1\over \pi} {\rm Im} \Pi^{\rm H}(t) dt
\end{equation}
which is a diagonal matrix
\begin{equation}
G^P_0 =\widetilde{\cal T}^aG_0^{\rm QCD}{\cal T}^a=
\begin{pmatrix}
{(G^P_0)}_{11} & 0 \\
0 & {(G^P_0)}_{22}
\end{pmatrix}
~.
\label{G_matrix}
\end{equation}

\begin{figure}[htb]
\centering
\includegraphics[width=\columnwidth]{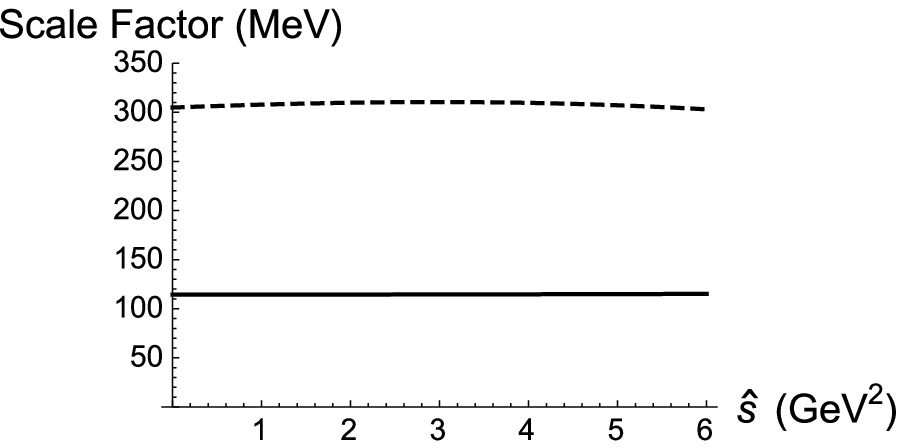}\hspace{0.05\columnwidth}
\caption{The scale factors $\Lambda$ (lower solid curve) and $\Lambda'$ (upper dashed curve) are shown as a function of $\hat s$
for optimized continuum thresholds $s_0^{(1)}=2.2\,{\rm GeV^2}$ and $s_0^{(2)}=4.9\,{\rm GeV^2}$.  The mixing angle $\cos\theta_a=0.493$ of Ref.~\cite{global} has been used.
}
\label{scale_fig1}
\end{figure}

As an illustration of how our proposed approach bridges chiral Lagrangian and QCD sum-rule methods we now calculate the scale factors $\Lambda$  and $\Lambda'$ for the isotriplet $a_0$ scalar meson system.  The QCD currents are \cite{Chen:2007xr,harnett_quark}
\begin{gather}
J^{QCD}=\begin{pmatrix}
J_1\\
J_2
\end{pmatrix}
\,,~
J_1=\left(\bar u u-\bar d d\right)/\sqrt{2}\\
\begin{split}
J_2&=\frac{\sin\phi}{\sqrt{2}}d^T_\alpha C\gamma_\mu\gamma_5 s_\beta\left(\bar d_\alpha\gamma^\mu\gamma_5 C\bar s_\beta^T-\alpha\leftrightarrow \beta \right)
\\
&+\frac{\cos\phi}{\sqrt{2}}d^T_\alpha C\gamma_\mu s_\beta\left(\bar d_\alpha\gamma^\mu C\bar s_\beta^T+\alpha\leftrightarrow \beta \right)
- u\leftrightarrow d
\end{split}
\end{gather}
where $C$ is the charge conjugation operator and $\cot\phi=1/\sqrt{2}$ \cite{Chen:2007xr}.
The diagonal terms in \eqref{G_matrix} give
\begin{gather}
G_{11}^H(\hat s,\tau)=a A
G_{11}^{QCD}\left(\hat s,\tau, s_0^{(1)}\right)-bB
G_{22}^{QCD}\left(\hat s,\tau, s_0^{(1)}\right)
\label{G_eqs}
\\
G_{22}^H(\hat s,\tau)=-aB
G_{11}^{QCD}\left(\hat s,\tau, s_0^{(2)}\right)+bA
G_{22}^{QCD}\left(\hat s,\tau, s_0^{(2)}\right)
\nonumber
\\
A=\frac{\cos^2\theta_a}{\cos^2\theta_a-\sin^2\theta_a}\,,~
B=\frac{\sin^2\theta_a}{\cos^2\theta_a-\sin^2\theta_a}
\\
a=\frac{m_q^2}{\Lambda^6}\,,~b=\frac{1}{\left(\Lambda'\right)^{10}}
\end{gather}
where the QCD continuum has been absorbed into the QCD Gaussian sum-rules,  $G_{11}^H$ and $G_{22}^H$ respectively represent $a_0(980)$ and  $a_0(1450)$, and the factor of $m_q^2$ is combined with  $G_{11}^{QCD}$  for renormalization-group purposes.  Note that each physical sum-rule has its own continuum threshold represented by $s_0^{(1)}$ and $s_0^{(2)}$.  Given an input of $\cos\theta_a$ from chiral Lagrangians \cite{global} and the physical mass and width of the $a_0$ states, one can solve  \eqref{G_eqs} for the (constant) scale factors $\Lambda$ and $\Lambda'$, and develop a procedure for the optimized values of the continuum thresholds that minimize the $\hat s$ dependence of the scale factors.  Expressions for the necessary QCD sum-rules can be found in Refs.~\cite{Chen:2007xr,GSR_qq_results_mix,harnett_quark} along with standard values of the QCD input parameters (e.g., QCD condensates). We choose $\tau=3\,{\rm GeV^4}$ consistent with the central value used in Refs.~\cite{GSR_qq_results_mix,harnett_quark}.  In Fig.~\ref{scale_fig1} we show the $\hat s$ dependence of the scale factors for the optimized values of the continuum.  The remarkable independence of the scale factors on the auxiliary sum-rule parameter $\hat s$ clearly demonstrates the validity of our proposed methodology.

The extracted values of the scale factors from Fig.~\ref{scale_fig1},  $\Lambda=115\,{\rm MeV}$ and  $\Lambda'=307\,{\rm MeV}$, are characteristic of the energy scales intuitively expected to emerge in relating QCD to chiral Lagrangian approaches.  However,  $\Lambda$ is also related to the vacuum expectation value
\begin{equation}
\langle S_1^1\rangle=-\frac{m_q\langle\bar u u\rangle}{\Lambda^3}=0.056\,{\rm GeV}\,,
\end{equation}
in remarkable agreement with the chiral Lagrangian prediction of $\langle S_1^1\rangle=0.061\,{\rm GeV}$ \cite{global}, providing strong supporting evidence for our proposed methodology.
Similarly,   $\Lambda'$  is related to the vacuum expectation value
\begin{equation}
\langle {S'}_1^1\rangle=1.31\frac{\langle \bar d d\rangle \langle \bar s s\rangle}{\Lambda'^5}\approx 0.04\,{\rm GeV}
\end{equation}
compared with the chiral Lagrangian prediction of $\langle {S'}_1^1\rangle\approx 0.03\,{\rm GeV}$ \cite{global}. Although this comparison is not expected to be as robust because it does not contain renormalization-group invariants,  the approximate agreement is significant.

We have also explored variations in $\cos\theta_a$ from the chiral-Lagrangian
value $\cos\theta_a=0.493$ \cite{global}; increasing to $\cos\theta_a=0.6$ and decreasing to $\cos\theta_a=0.4$ leads to clear increases in the $\chi^2$ measure used to optimize the predicted scale factors. Thus the QCD content of our proposed methodology has sufficient sensitivity to  discriminate between different chiral Lagrangian mixing scenarios.
As an additional consistency check, we have verified that  Eq.~\eqref{G12} for the off-diagonal sum-rule $G_{12}^{QCD}$ agrees with the estimated order-of-magnitude effects expected from the leading-order perturbative result.

In summary, we have proposed a general methodology that connects chiral Lagrangians to QCD sum-rules through  scale factor matrices relating the two approaches.
These scale factor matrices combined with a rotation matrix then provide a matrix used to obtain the physical projection of the QCD correlation function matrix onto mesonic states.
 A detailed implementation has been provided for the $a_0(980)$-$a_0(1450)$ system, and the resulting QCD extraction of the scale factors are independent of the auxiliary sum-rule parameter and are  in excellent agreement with the quark QCD condensates and vacuum expectation values in chiral Lagrangians.  In this example implementation, we find sufficient sensitivity to discriminate between different chiral Lagrangian mixing angles.  Thus our proposed methodology provides a powerful bridge connecting the relevant low-energy chiral Lagrangian models and experimental data to the theoretical framework of QCD. We expect that this powerful synergy between chiral Lagrangians and QCD will permit future progress on  more challenging and controversial aspects of low-energy hadronic physics.

\section*{Acknowledgments}
TGS is grateful for the hospitality of AHF, Dr.~A.~Wolfe and SUNY Polytechnic Institute as well as partial funding from UUP Individual Development Award (2014, 2015) and President Opportunity Fund (2014), while this work was initiated and completed.
TGS and AP are partially supported by research funding from the Natural Sciences and Engineering Research Council of Canada (NSERC).

\end{document}